# Mapping the Initiation of Plastic Deformation in Nanoindentation

D.J. Dunstan[1 *], T.T. Zhu[1], M. Hopkinson[2] and A.J. Bushby[1]

[1]Centre for Materials Research, Queen Mary University of London,

London E1 4NS,

[2]Department of Electronic and Electrical Engineering, University of Sheffield,

Mappin Street, Sheffield S1 3JD.

**\* Contact author:**

Prof. D.J. Dunstan,

Physics Department, Queen Mary University of London, London E1 4NS.

Tel:  020 7882 3687

Fax : 020 8981 9465

e-mail:  d.dunstan@qmul.ac.uk

**Abstract**

Under the inhomogenous stress field set up by nanoindentation, the smaller the extent of the stress field the greater the yield pressure. When the specimen contains a thin layer of softer material, the yield pressure is reduced if plasticity initiates in the soft material. A series of specimens with the soft layer at different depths enables the depth at which plasticity initiates to be mapped. InGaAs lattice-matched to InP is used, and the soft layer is 320 nm of strained-layer InGaAs superlattice. Under nanoindentation, we find that plastic yield initiates throughout a region ranging from 300nm to over 1μm according to the indenter tip radius. The region matches the depth range over which the stress exceeds the yield stress of the InGaAs. Thus a requirement for yield is an overload, or excess stress, throughout a finite volume.

In micromechanics, materials and structures are found to be stronger when plastic yield by dislocations is restricted to small volumes; this is called the size effect. It is observed in many systems and increasingly exploited as technology moves to smaller scales. Explanations are controversial, including strain gradient [1–3], dislocation starvation [4], statistical distributions of dislocation sources [5,6] and the critical thickness effect [7–12]. We have previously proposed that plastic yield begins over a finite volume and that this itself is the cause of increased yield strength whenever plasticity is restricted to small volumes [13]. Here, we map this volume by indenting purpose-built semiconductor crystal structures. We find that it matches the region over which stress exceeds bulk yield stress. This shows that classical yield criteria for plasticity do indeed apply, but demands the additional requirement that yield criteria are met throughout a finite volume, leading to overstress in all or part of the volume. This explains unambiguously the size effect at yield.

According to the classical Hertzian theory, the maximum resolved shear stress occurs on the axis of the indenter and at a depth of about ½$a$ beneath the surface, where $a$ is the radius of the projected area of contact [14]. This position is the point of initial yield according to the usual yield criteria such as the Mises or Tresca criteria [14]. The design of the experiment is shown schematically in Fig.1. If a solid specimen of a hard material contains a thin layer of slightly softer material, the yield pressure should be reduced when the zone of initial plasticity is within the soft layer. Putting the soft layer at different depths will map the zone by the reduction in the yield pressure when the soft layer intersects the zone. InGaAs alloy strained-layer superlattices have a yield pressure in indentation significantly lower than the yield pressure of bulk InGaAs with the same average composition [10,15,16]. They can be incorporated into bulk unstrained InGaAs single-crystal material grown on InP substrates. Only the composition (and therefore the coherency strain) changes slightly across the interfaces. Other properties such as elastic moduli scarcely change [17]. We grew a series of samples with a thin soft layer of superlattice at various depths within harder bulk InGaAs. For transmission electron micrographs of such structures see Fig.9 of Ref. 15. Control specimens of bulk unstrained InGaAs with no buried layer, and of bulk superlattice, were also grown. Details of the design and growth are given in the Supplementary Online Materials.

To display the size effect and to map any variation of the initial plastic zone with indenter radius, indentation was performed with four spherical indenters with radii $R$ from 0.45μm to 4.8μm. The multiple partial unloading technique [18] was used and the mean pressure at the onset of gross plasticity was identified as described in Zhu *et al.*[19]. The indentation analysis gives a contact radius $a$ at yield and a mean pressure at yield $P_Y$ for each indent [18,19], and the results plotted in Fig.2 are averages over 49 indents for each datum. It is rather important that our analysis avoids the issue of 'pop-in', i.e. of elastic overload due to a paucity of dislocation sources. As shown by Bei *et al.* [5] and Morris *et al.* [6], there is a statistical size effect in which shortage of sources gives yield points ranging up towards the theoretical strength. Our analysis gives the lower limit of this statistical distribution, i.e. the yield strength when there is no source constraint [19]. This lower limit of the statistical distribution of $P_Y$ varies with indenter radius – that is the size effect.

The data are plotted on the left-hand graphs of Fig.2. Note that the *x*-axes of these four graphs, for the four indenters used, are different, showing clearly the indentation size effect. Consider Fig.2(a) first, for the largest indenter. The yield pressures of the superlattice (SL) control specimen and the bulk InGaAs control specimen are marked by heavy lines, and the data points are the yield pressures of the buried layer specimens plotted against the centre depths of the buried layers. The shallowest soft layer (at the surface) has little effect, the yield pressure being close to that of the bulk InGaAs control value. Full plastic flow in indentation is commonly considered to take place in a hemisphere of radius ~$a$ centred at ($r = 0$, $z = 0$) [20]. While this is a good approximation for larger indentation strains, it cannot be the case for the onset of yield detected here, for the soft layer at the surface would have as great an effect as any deeper soft layer – which is clearly not true of the data in Fig.2(a). The data suggest that yield initiates sub-surface, as expected from classical theory [14]. The maximum effect of the soft layer at greater depths reduces the yield pressure only about half-way to the superlattice control value. This suggests immediately that the plastic yield cannot be initiated at the point where the classical yield criterion is first met. Instead, a range of depths much larger than the thickness of the soft layer must be implicated together. That is, classical yield criteria alone do not work here.

The reduction in yield pressure is expected to be proportional to the fraction of the zone of initial plasticity overlapping the soft layer. The thickness of the soft layers corresponds to a top-hat function. We approximate the zone of initial plasticity as a sphere. The effect of the soft layer as a function of its depth, $z_L$, is then proportional to the fraction of the volume of the sphere which overlaps the top-hat function. The calculation of this function, $F(z_L)$, is straightforward (see Supplementary On-line Information). It has a maximum when the sphere and the soft layer are centred at the same depth. It is zero when the sphere and the layer do not overlap, and it has the maximum value of unity if the sphere lies wholly within the top-hat. A least-squares fit of the function $F(z)$ to the data, with $F = 1$ at the bulk superlattice control yield pressure and $F = 0$ at the bulk InGaAs control yield pressure, made with the depth and radius of the sphere as free fitting parameters, is shown as the dashed red curve in Fig.2(a). Note that the amplitude of the function $F$ is not itself a fitting parameter. It is determined by the experimental yield pressures for the bulk InGaAs and bulk superlattice control samples, the known thickness of the soft layer and the fitted radius of the sphere, and it agrees well with the data.

In the right of Fig.2(a) we compare the fitted sphere with the information we have from the theory of spherical indentation [14]. The depth and radius of the sphere are shown schematically by the red dashed semi-ellipse. The shear stress $\tau_{rz}(z)$ on-axis at yield is calculated from Hertzian mechanics [14] with the contact radius $a$ and the yield pressure $P_Y$ as parameters (see Supplementary On-line Material), and plotted as a function of depth $z$. The solid green line shows the depth profile (the top-hat function) of the yield strength in shear of one of the five specimens, corresponding to the circled data-point, using the bulk and superlattice control values of $\tau_Y$ extrapolated to large indenters ($1/R \rightarrow 0$; see Supplementary On-line Material). The shear yield strength may also be considered to be the shear flow stress for small flow, and we are directly assuming that it does not depend on size. The region in which the shear stress exceeds the shear yield strength of bulk InGaAs defines an overloaded region and the excess stress is shaded blue. Remarkably, the upper and lower bounds of the sphere of initial plasticity of the fitted sphere (dashed red semi-ellipse) are close to the upper and lower limits of the overloaded region. Consequently, we model the effect of a sphere of initial plasticity which matches precisely the overloaded volume, with the depth and radius reported in Table S2 (Supplementary online information), indicated schematically by the solid blue semi-ellipse. Calculating the function $F(z_L)$ for the model sphere, with no fitting parameters, gives the solid blue curve compared with the data in the left of Fig.2(a). Agreement with the data is good.

Fig.2(b) repeats (a) but for the 3.4μm indenter. Again the model fit to the data (solid blue curve) is good, comparable with the least squares fit (dashed red curve). For the two smallest indenters, only two of the buried-layer specimens reduced the yield pressure below the bulk InGaAs value (Fig.2(c) and (d)). This is insufficient data to permit a valid least-squares fit. Here we give only the fits assuming the model sphere delimited by the overloaded volumes shaded blue in Fig2.(c) and (d). The fit to the data is still satisfactory, and it is worth noting that for these small indenters the function $F(z_L)$ peaks at unity. That is, when the zone of initial plasticity is entirely within the soft layer, it is indeed the yield pressure of the soft layer that is observed. It follows, as a central conclusion of this paper, that for the two larger indenters the departure from the elastic line is initiated by conditions over an extended zone larger than the thickness of the buried layers. Moreover, it is then the asymmetry of the shear stress $\tau_{rz}(z)$ function about its peak at $z = ½a$ which explains why the least-squares fitted values of $z_L$ – i.e. the depths at which the soft layer has maximum effect – are significantly greater than ½a, as seen in Fig.2.

The quality of fit to the data shows that initial yield depends on conditions throughout a volume up to micron size and that a sphere is not a bad approximation for the shape of this volume. The soft layer has essentially the same elastic properties as the bulk crystal so its presence modifies only the plastic behaviour. It follows that for it to affect the plastic yield pressure, it must itself yield. Thus agreement between the two methods, i.e. the direct imaging of initial plasticity (the least-squares fit to the data) and the determination of the overloaded region by the points where the stress crosses the yield stress, can only happen if plasticity initiates simultaneously throughout the overloaded volume.

The remarkable agreement observed shows that there are two criteria which must be satisfied before plastic deformation occurs. Firstly, as usual, the stress must attain the yield stress. Secondly, the stress must exceed the yield stress over an overloaded volume which is a region of finite size. A more sharply peaked stress function, as generated under the smaller indenters, generates a higher peak stress within a smaller overloaded volume as seen in Fig.2.

This is an important result because it shows that the initiation of plastic deformation is raised to a higher applied stress simply by a restriction of the volume of material under stress. Explicitly, a restriction of the volume requires an overstress. The yield strength of the material itself is not changed but size itself creates the size effect. This is an explanation of the size effect which is consistent with critical thickness theory [7–12], but quantitative agreement remains to be demonstrated. However, the data presented here represent a significant challenge to other theories of the size effect.

**Acknowledgements**

We are grateful to Dr W.J. Clegg and Dr S. Korte, Cambridge, and to Dr A.J. Drew, Queen Mary, for many useful discussions about these results and for critical readings of the manuscript. We acknowledge the UK Engineering and Physical Sciences Research Council for financial support for this programme (EP/C518004).

**References**

[1] N.A. Fleck, G.M. Muller, M.F. Ashby and J.W. Hutchinson, Acta Met. **42**, 475 (1994).

[2] J.S. Stölken and A.G. Evans, Acta Mater. **46**, 5109 (1998).

[3] M.I. Idiart, V.S. Deshpande, N.A. Fleck and J.R. Willis, Int. J. Eng. Sci. **47**, 1251 (2009).

[4] J.R. Greer and W.D. Nix, Phys. Rev. B**73**, 245410 (2006).

[5] H. Bei, S. Shim, G.M. Pharr and E.P. George, Acta Mat. **56,** 4762 (2008).

[6] J.R. Morris, H. Bei, G.M. Pharr and E.P. George, MRS Fall Conference, Boston, GG2.7, (2009).

[7] J.W. Matthews and A.E. Blakeslee, J. Crystal Growth **27,** 118 (1974).

[8] E.A. Fitzgerald, Mater. Sci. Rep. **7**, 87 (1991).

[9] D.J. Dunstan, J. Mater. Sci.: Materials in Electronics **8**, 337 (1997).

[10] N.B. Jayaweera, J.R. Downes, M.D. Frogley, M. Hopkinson, A.J. Bushby, P. Kidd, A. Kelly and D.J. Dunstan, Proc. R. Soc. Lond. **A459,** 2049 (2003).

[11] B. Ehrler, X.D. Hou, T.T. Zhu, K.M.Y. P'ng, C.J. Walker, A.J. Bushby and D.J. Dunstan, Phil. Mag. **88**, 3043 (2008).

[12] D.J. Dunstan, B. Ehrler, R. Bossis, S. Joly, K.M.Y. P'ng and A.J. Bushby, Phys. Rev. Lett. **103**, 155501 (2009).

[13] D.J. Dunstan and A.J. Bushby, Proc. R. Soc. Lond. **A460,** 2781 (2004).

[14] K.L. Johnson, *Contact Mechanics*. Cambridge University Press, Cambridge, 1985).

[15] S.J. Lloyd, K.M.Y. P'ng, W.J. Clegg, A.J. Bushby and D.J. Dunstan, Philos. Mag. **85,** 2469 (2005).

[16] S. Korte and W.J. Clegg, Acta Mat. **58,** 59 (2009).

[17] M.E. Brenchley, M. Hopkinson, A. Kelly, P. Kidd and D.J. Dunstan, Phys. Rev. Lett. **78**, 3912 (1997).

[18] J.S. Field and M.V. Swain, J. Mater. Res. **8,** 297 (1993).

[19] T.T. Zhu, A.J. Bushby and D.J. Dunstan, J. Mech. Phys. Solids **56,** 1170 (2008).

[20] D. Tabor, Proc. Roy. Soc. (London) A**192**, 247 (1948).

**Figure Captions**

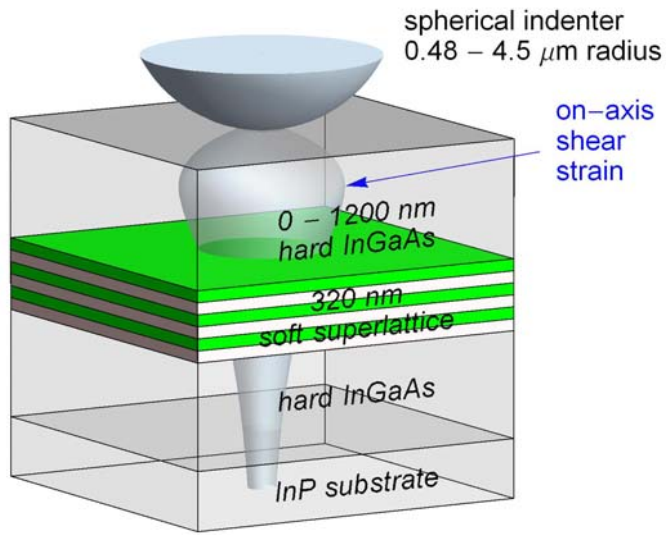

**Fig.1:** (Color Online) Schematic of the experiment. A bulk InGaAs layer has a thin strained-layer superlattice incorporated at various depths. Under a spherical indenter, the shear stress on-axis under the indenter peaks below the surface, as indicated by the diameter of the gourd-shaped object below the indenter, and plastic yield is expected to initiate where the shear stress is greatest. When the soft superlattice intersects this volume, the yield pressure is reduced.

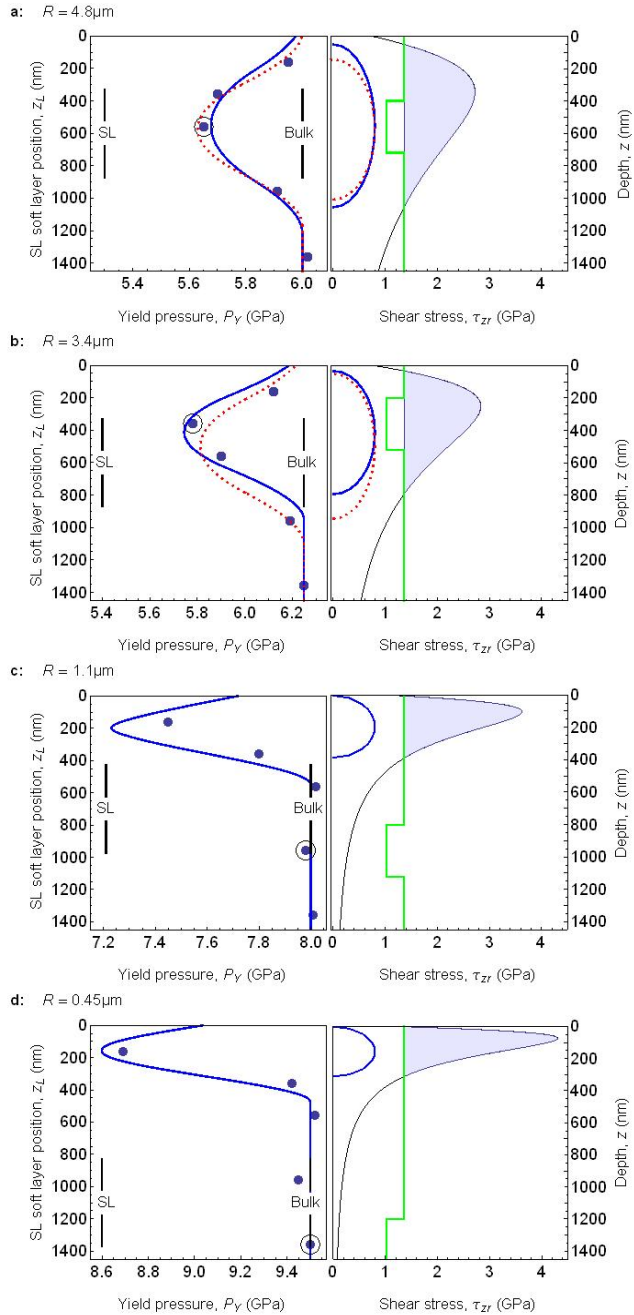

**Fig.2:** (Color Online) Yield pressure data and schematics of their interpretation. The data for the 4.8μm radius indenter are plotted on the left graph of Panel (a) against the depth of the soft layers. The dashed red curve is the least-square fit of the function $F(z_0)$. On the right graph, the sphere obtained from the least-squares fit is indicated by the dashed red semi-ellipse. The shear stress on axis is plotted as a function of depth. The solid green line shows the yield strengths in shear, drawn as a function of depth for one particular specimen (corresponding to the ringed data point in the left graph). The shaded region is the overloaded volume. The blue semi-ellipse shows the sphere that matches the overloaded volume. The solid curve in the left graph shows $F(z_0)$ calculated for this sphere. Panel (b) repeats (a) but for the 3.4μm indenter. The circled data point and the solid green line are for a different specimen. Panels (c) for the 1.1μm and (d) for the 0.45μm indenters are the same except that no least-squares fits are given.

# Supplementary Material
Dunstan *et al.*, "*Mapping the initiatiation of. . .*"

## 1. Methods

The samples were $In_xGa_{1-x}As$ alloy structures grown on (001)-oriented InP substrates by molecular beam epitaxy at low growth temperatures ~ 420°C to prevent phase decomposition of the InGaAs alloy. Further details of the growth and characterisation have been given previously.[10,22] Each sample is a lattice-matched layer of $In_{.53}Ga_{.47}As$ 2.5 μm thick within which is included a strain-balanced $In_xGa_{1-x}As$ superlattice 320nm thick positioned at different depths. In the specimens used here, the alternate tensile and compressive layers of the superlattice material were each 20nm thick, and of nominal compositions $x = 0.53 \pm 0.12$, giving strains of $\pm 0.008$. The layer compositions were confirmed by X-ray diffraction and simulation analysis[22] to be within $\Delta x \sim \pm 0.01$ of the nominal values.

Indentation was performed using a UMIS 2000 nano-indentation instrument using a load-unload protocol.[23] Four diamond indenters were calibrated using sapphire, silica and GaAs standards to give effective radius as a function of depth.[24] A 7×7 array of indents is specified. For each indenter the maximum load is specified (about 30% above the expected load required for yield, i.e. $R$ = 0.45 μm, 8mN; 1.1 μm, 12mN; 3.4 μm, 50mN; 4.8 μm, 100mN) and reached in sixty increments. After each load increment, the load is reduced to 75% (unload). The penetration depths at load and unload are recorded and used in the analysis. Data from indents showing significant pop-in is discarded and replaced by data by second and if necessary subsequent 7×7 arrays of indents made under the same conditions, so that each $P_Y$ reported is an average over 49 indents without significant pop-in.

## 1. InGaAs Growth and Superlattice Design

From previous work,[10] it was known that symmetrical strain-balanced superlattices display a lower yield pressure than bulk material, in proportion to the strain and independent of the period of the superlattices. It was known that contact radii for the smallest indenter might be in the region of 150nm. Choosing therefore a thickness of about 300nm for the soft layers placed the depth of ½$a$ firmly within the shallowest soft layer for the smallest indenter and well outside the shallowest soft layer for the largest indenter. Within such a small thickness, and without a clear understanding of what makes the superlattices soft, it was desirable to use thinner superlattice repeat periods than the standard 100nm in Ref.10 to ensure that bulk superlattice behaviour would be obtained. 20nm layer thicknesses provide this without dropping below the previous thinnest layers of 17nm. While compressive and tensile strains up to 0.015 can be grown, the moderate mismatch strain of 0.008 was chosen as a compromise between maximising the effect, and not prejudicing growth quality throughout the rather large 2.5μm total epitaxial growth thickness.

## 2. Indentation and Yield Pressure

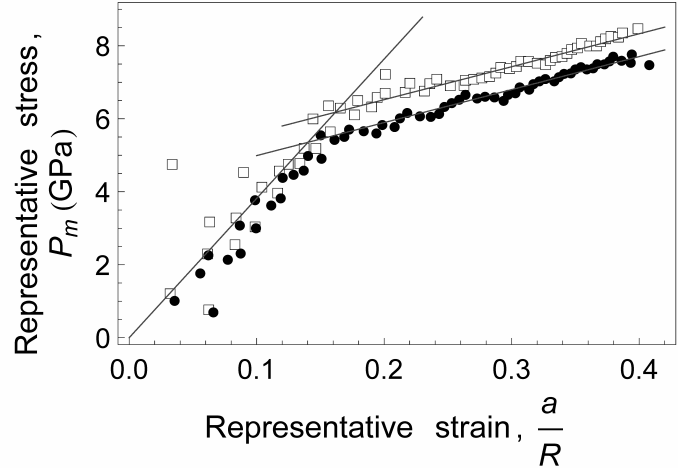

**Fig.S1:** Indentation data for single indents with a 3.4μm radius indenter into the bulk unstrained InGaAs (open squares) and the bulk superlattice (solid circles) control specimens are shown. The lower straight line is the theoretical elastic curve. The upper lines are linear least squares fits to the data in the plastic regime. The intercepts with the elastic line give the yield pressures, $P_Y$.

Data for single indents using the indenter of radius $R = 3.4$ μm in the bulk InGaAs and bulk superlattice control samples are plotted in Fig.S1. The raw data (force, penetration depth and calibrated indenter radius) are used to calculate the contact radius $a$, hence the contact area and the mean pressure $P_m$ on the contact area using Hertzian contact mechanics[23,24]. $P_m$ is taken to represent stress in indentation, and strain is represented by $a/R$. According to Zhu et al.,[23] the best estimate of the yield pressure, $P_Y$, is obtained for indents both with and without pop-in by fitting a straight line to the data above yield and taking the intercept of this line with the theoretical elastic line. In Fig.S1, the elastic line is $y = 38.19x$ and the least-square fits to the data above yield are $y = 9.026x + 4.719$ (bulk InGaAs) and $9.046x + 4.086$ (superlattice), giving intercepts with the elastic line at $y = 6.18$ GPa and 5.35 GPa respectively (compare Table I). The values given in Table I are obtained by averaging over 49 indents, after discarding the data for indents with large pop-in.

The data in Table I show the size effect clearly, as an increase in $P_Y$ for smaller indenters. As previously shown[23] the increase of $P_Y$ is linear within error with $R^{-1/3}$ and with $a^{-1/2}$. Extrapolations of the data for the control specimens to $R^{-1/3} = 0$ and to $a^{-1/2} = 0$ give yield pressures for infinitely large indenters (i.e. with no size effect) on the control InGaAs and the control superlattice specimens (Fig.S2). Least-squares linear fits in both plots agree in giving the yield pressures for large indenters as 2.99 GPa and 2.99 GPa for the InGaAs, and 2.35 GPa and 2.31 GPa for the superlattice (Table I and Fig.3).

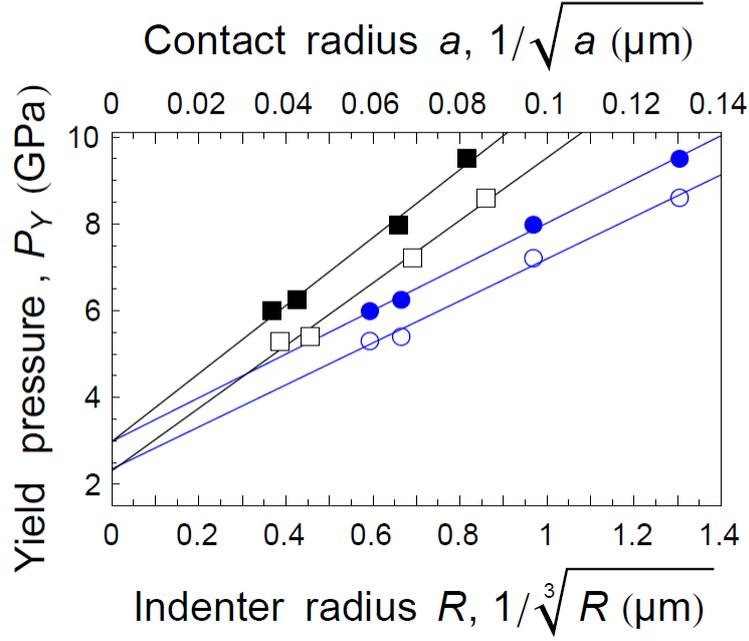

**Fig.S2:** The mean pressures at yield $P_Y$ are plotted for the bulk InGaAs sample (solid data points) and for the thick superlattice sample (open data points), against the inverse square root of the contact area $a$ (black squares, upper x-axis) and against the inverse cube root of the indenter radius $R$ (blue circles, lower x-axis). The solid lines are least-squares fits of $y = mx + c$ to the data.

## 2. Calculation of Shear Stresses

From the Hertzian theory of spherical indentation,[16] the peak pressure at the centre of the indent is $P_0 = 3/2\ P_m$, The radial and azimuthal stresses $\sigma_r$ and $\sigma_z$ as functions of depth $z$ on-axis below the indenter are

$$\sigma_r = -\frac{4}{3}P_0\left(\left(1-\frac{z}{a}\tan^{-1}\frac{a}{z}\right)+\frac{1}{2(1+z^2/a^2)}\right)$$

$$\sigma_z = -\frac{4}{3}P_0\left(\frac{1}{1+z^2/a^2}\right)$$

The shear stress on axis is $\tau_{rz} = \frac{1}{2}(\sigma_r - \sigma_z)$ and its peak value, at a depth of $\sim\frac{1}{2}a$ is $\tau_{max} = 0.31\ P_0$. Hence, in the absence of the size effect, the yield stresses in shear for the bulk InGaAs ($P_m$ at yield of 3.0 GPa) and for the bulk InGaAs ($P_m$ at yield of 2.3 GPa) are $\tau_Y = 0.45 P_m$, or 1.36 GPa and 1.04 GPa respectively. These are the values plotted as the solid green lines in Fig.1, (b, d, f, h).

## 4. Calculation of F(z)

The convolution or overlap integral between the top-hat function of the yield strength as a function of centre depth $z_L$ and thickness $h$, and the sphere of initial plasticity, centre depth $z_S$ and radius $\rho$, may be calculated as follows. Let the origin of depth $z$ be at the centre of the sphere ($z_S = 0$) and write $P_Y(z)$ as

$$P_Y = 0 \quad |z - z_L| > h/2$$
$$P_Y = 1 \quad |z - z_L| < h/2$$

Then for $2\rho > h$, and assuming the free surface is above the top of the sphere,

$$F(z_L) = 0 \qquad\qquad z_L < -\rho - h/2$$
$$= \int_{-\rho}^{z_L+h/2} \pi(\rho^2 - z^2)dz \qquad -\rho - h/2 < z_L < -\rho + h/2$$
$$= \int_{z_L-h/2}^{z_L+h/2} \pi(\rho^2 - z^2)dz \qquad -\rho + h/2 < z_L < \rho - h/2$$
$$= \int_{z_L-h/2}^{\rho} \pi(\rho^2 - z^2)dz \qquad \rho - h/2 < z_L < \rho + h/2$$
$$= 0 \qquad\qquad z_L > \rho + h/2$$

Similarly, for $2\rho < h$, and again assuming the free surface is above the top of the sphere,

$$F(z_L) = 0 \qquad\qquad z_L < -\rho - h/2$$
$$= \int_{-\rho}^{z_L+h/2} \pi(\rho^2 - z^2)dz \qquad -\rho - h/2 < z_L < -\rho + h/2$$
$$= 1 \qquad\qquad -\rho + h/2 < z_L < \rho - h/2$$
$$= \int_{z_L-h/2}^{\rho} \pi(\rho^2 - z^2)dz \qquad \rho - h/2 < z_L < \rho + h/2$$
$$= 0 \qquad\qquad z_L > \rho + h/2$$

It is then straightforward to shift the origin of the function for a sphere centred at $z_S$, and to rescale the limiting values of the function, 0 to the bulk InGaAs $P_Y$ and 1 to the superlattice $P_Y$ to obtain the function $F(z)$ plotted in Fig.1 (a, b, c, d). Least-squares fitting to the data for the two largest indenters gives the values for the sphere centre depth and radius given in Table S2. Values for all four indenter sizes, obtained by identifiying this zone with the overloaded volume, are also given in Table S2.

**Table S1:** Indentation data. Pressures $P_Y$ and contact radii $a$ at yield.

| Indenter radius $R$ | | Bulk InGaAs | 2.5μm SL | Buried layer centre depth (nm) | | | | |
|---|---|---|---|---|---|---|---|---|
| | | | | 160 | 360 | 560 | 960 | 1360 |
| $R \to \infty$ | $P_Y$ GPa | 3.0 | 2.3 | - | - | - | - | - |
| 4.8μm | $P_Y$ GPa | 5.99 | 5.30 | 5.95 | 5.70 | 5.65 | 5.91 | 6.02 |
| | $a$ nm | 740 | 670 | 745 | 710 | 700 | 740 | 740 |
| 3.4μm | $P_Y$ GPa | 6.25 | 5.40 | 6.12 | 5.78 | 5.90 | 6.19 | 6.25 |
| | $a$ nm | 550 | 480 | 540 | 515 | 525 | 550 | 550 |
| 1.1μm | $P_Y$ GPa | 7.98 | 7.21 | 7.45 | 7.80 | 8.02 | 7.98 | 8.01 |
| | $a$ nm | 230 | 210 | 215 | 225 | 230 | 230 | 230 |
| 0.45μm | $P_Y$ GPa | 9.50 | 8.60 | 8.69 | 9.42 | 9.52 | 9.45 | 9.50 |
| | $a$ nm | 150 | 135 | 135 | 150 | 150 | 150 | 150 |

**Table S2:** Radius $\rho$ and centre depth $z_S$ of the sphere of initial plasticity

| Indenter radius | Least-squares fit | | Theoretical fit | |
|---|---|---|---|---|
| | $z_S$ nm | $\rho$ nm | $z_S$ nm | $\rho$ nm |
| 4.8μm | 574 | 432 | 553 | 502 |
| 3.4μm | 498 | 447 | 413 | 380 |
| 1.1μm | - | - | 198 | 192 |
| 0.45μm | - | - | 147 | 145 |